\documentclass[11pt]{article}

\usepackage[]{graphicx}
\usepackage{amsmath}
\usepackage{amssymb}
\usepackage{epsfig}

\usepackage{graphicx}
\begin{document}


\begin{titlepage}

\renewcommand{\thefootnote}{\alph{footnote}}
\vspace*{-3.cm}
\begin{flushright}

\end{flushright}

\vspace*{0.3in}

\renewcommand{\thefootnote}{\fnsymbol{footnote}}
\setcounter{footnote}{-1}

{\begin{center} {\Large\bf  Exact two-point resistance, and the simple random walk on the complete graph minus  $ N$  edges  }
\end{center}}
\renewcommand{\thefootnote}{\alph{footnote}}

\vspace*{.8cm} {\begin{center} {\large{\sc
                }}
\end{center}}
\vspace*{0cm} {\it
\begin{center}
 \vspace*{.8cm} {\begin{center} {\large{\sc
                Noureddine~Chair
                }}
\end{center}}
\vspace*{0cm} {\it
\begin{center}
 Physics Department,
 University of Jordan, Amman, Jordan

Email: n.chair@ju.edu.jo
\end{center} }
\end{center} }

\vspace*{1.5cm}

\begin{center} Abstract\end{center} An analytical approach is developed  to obtain the exact  expressions for the  two-point resistance, and the total effective resistance of the complete graph minus $N$ edges of the opposite vertices. These expressions are written in terms of certain numbers that we introduced which we call the Bejaia and the Pisa numbers, these numbers are the natural generalizations of the bisected  Fibonacci and Lucas numbers.  The correspondence between random walks and the resistor networks is then used to obtain the exact  expressions for the  the first passage and mean first passage times on this graph.
 \vspace*{.5cm}
\end{titlepage}

\renewcommand{\thefootnote}{\arabic{footnote}}
\setcounter{footnote}{0}
\section{Introduction} 
A random walk on an undirected connected graph
G is a process that starts at some vertex of $G$, and
at each time step moves to one of the neighbors of the current vertex, each of them chosen with equal probability. The basic quantity relevant to random walks is the first passage time ( FPT ), or the hitting time,  this is the expected time to hit a target node for the first time  for a walker starting from a source node. This quantity is an indicator that characterize the transport efficiency and carries much information of random walks. It has been shown that the escape probability, the (FPT) and the commute time (the random round tripe between two nodes) of  random walks are related to the effective resistance \cite{Doyle, Tetali, Chandra}. Therefore, the effective resistance provides an alternative way to compute the (FPT).  A nice interpretation of the two-point resistance  $R_{ij}$  between  nodes   $i$ and $ j$  was given  by Klein and Randic \cite{randic}, as a measure of how close these  nodes are: for unit conductances,  $R_{ij}$  is small when there are many paths between the nodes $i$ and $ j$ , and large when there are few paths, between the  nodes $i$ and $ j$. With this interpretation in mind, the two-point  resistance sometimes is called  the resistance distance between nodes $i$ and $ j$, i.e., the two-point resistance enjoys the properties  of a distance function. An interesting quantity related to the two-point resistance in a resistor network (undirected  graph $G=(V,E)$ with vertex set $V$  and edge set $E$, with unit resistors as edges) is the total resistance distances of the graph $G$, denoted by  $ R(G)$. Recently, this quantity was shown to be equal to the network criticality \cite{Garcia}: a measure for robustness of a network to changes in traffic, topology, and community of interest.  The computation of the two-point resistance of graphs are usually difficult to obtain in a closed form, however, for  certain graphs with symmetries like the undirected circulant graphs, this may be possible. The undirected circulant graph \cite{Biggs}, is a graph   whose vertices can be ordered so that the adjacency matrix is a symmetric  circulant matrix, the $N$-cycle  and the complete graphs are examples of the circulant graphs.  Chau and Basu  \cite{ Chau}, derived recently  a formula to compute the (FPT) of the random walk on the $N$-cycle graph with $2p$ neighbors, i.e., the undirected circulant graph of the type $ C_N(1,2,\cdots,p)$ . Their formula is based on Lovasz's  formula for the expected hitting time of random walk on a finite graph \cite{Lovasz}.  Wu in his paper \cite{wu}, on the theory of resistor network,  derived a formula to compute the two-point resistance between any two nodes in terms of the eigenvalues and the eigenvectors of the Laplacian matrix associated with the finite electrical network. By  using his formula,  he obtained  two-point resistance of  the complete and  the cycle graphs. By diagonalizing the Laplacian matrix associated  with the $N$-cycle graph with $2p$  nearest neighbors and using Wu's formula we obtain a formula to compute the  two-point resistance between any two vertices of this  graph. Then it is not difficult to show that this formula  when multiplied by the number of  edges $\vert E\vert  $  is identical to the (FPT) given in \cite{ Chau}. This is expected, since we are dealing with the undirected circulant graphs that enjoy rotational symmetry,  each vertex of these  graphs  is a vertex-transitive, i.e., ‘looks the same’ from any vertex, then  the first passage time is symmetric under the exchange of the vertices. Therefore, we may as well consider that the random walk has  started at vertex $0$,  and after some steps reaches a given vertex say $l$. Using the commute time  formula given by  Chandra et al.  \cite{Chandra}, $ C_{ij }=2 \vert E\vert R_{ij}$, then  the first passage time  $H_{0l }$ may be written as  $H_{0,l }=\vert E\vert R_{0,l} $. For example, the two-point resistance between the vertex $0$ and any other vertex $l$ of the $N$-cycle  is $ R_{0,l}= l(1-l/N)$, and since  the number of edges is $ \vert E\vert=N$, then the expression for the (FPT) of the random walk on the $N$-cycle gives   $H_{0,l }=l(N-l)$. This result  was  derived previously using probabilistic techniques  on graphs \cite{Aldous}. In this paper, we give the  exact expression for the  two-point resistance between any two vertices of the of the  complete graph minus $N$ edges of the opposite vertices, $N$ is assumed to be odd, this graph is denoted by $ K_{N}^{-N}$. If $N$ is even, then, the complete graph minus $N/2$ edges of the opposite vertices is known as the cocktail-party graph \cite{Biggs}, in this case the the two-point resistance  computations are straightforward unlike in this paper. The general formula to compute the two-point resistance of the  graph   $ K_{N}^{-N}$ turns out to be given by trigonometrical power sums. To obtain the exact two-point resistance, an extra care is needed to use a formula by Schwatt \cite{ Schwatt} on trigonometrical power sums, since the latter does not give the right answer when the powers are congruent to $N$. Therefore, we have to solve this problem first before  doing our  computations. As a consequence,  computing  the two-point resistance is not direct and is done in steps, once  the right formula for the trigonometrical power sum is obtained,  we use the binomial coefficients representation by residues, and the linearity property of the residue operator. This property  played  an important role in our paper, that enabled us  to  avoid  carrying out certain  sum of binomials like  $\sum_{p=1}^{[j/N]}(-1)^{pN}\binom {2j}{j-pN} $,  this turns out to be  an open mathematical  problem in combinatorics $^{}$\footnote{The author would like to thank W. H. Gould and R. Sprugnol for  correspondence on this problem.}, the only known closed formula  for this sum is for $N=1,2,3$. Then, using  the Chebyshev polynomial of the first kind, and introducing certain numbers which we call the Bejaia and the Pisa numbers, the two-point resistance is obtained. We find that the names Bejaia and Pisa fit nicely, here, simply because Fibonacci started thinking about his famous numbers while he was in bejaia and wrote them when he went back to Pisa.  These numbers,  have nice properties like the Fibonacci and the Lucas numbers. More precisely, these numbers are the  natural generalizations of the bisected Fibonacci and Lucas numbers, that is, $ F_{2n}=\frac{1}{\sqrt{5}}\Bigg(\frac{3+\sqrt{5}}{2}\bigg)^{n}-\frac{1}{\sqrt{5}}\Bigg(\frac{3-\sqrt{5}}{2}\bigg)^{n}$, and  $ L_{2n}=\Bigg(\frac{3+\sqrt{5}}{2}\bigg)^{n}+\Bigg(\frac{3-\sqrt{5}}{2}\bigg)^{n}$respectively. The total effective resistance, and the important parameters of random walks on this graph such as the (FPT), and the mean first  passage times (MFPT) are also given by exact expressions.  It is interesting to note that for regular graphs,  we obtained the (MFPT) expression  through a very simple formula in terms of  The total effective resistance, and the degree of the graph..
\section{The two-point resistance  of the complete graph minus  $N$ edges}
In this section, exact formula for the  two-point resistance of the  complete graph minus $N$ edges of the opposite vertices is obtained, here, $N$ is assumed to be odd, this graph is denoted by $ K_{N}^{-N}$. Our computations  are done in steps and somehow similar to the computations of the two-point resistance of the $N$-cycle with four nearest neighbors carried out recently by the author \cite{Chair1}. Our computations  of the two-point resistance  is based on a theorem by Wu \cite{wu},  which  states  that for a  resistor network with unit resistance,  the the two point resistance between any two nodes $\alpha$ and $\beta$ is given by 
\begin{equation}
\label{7}
R_{\alpha,\beta}=\sum_{n=1}^{N-1}\frac{|\psi_{n\alpha}-\psi_{n\beta}|^2}{\lambda_n},
\end{equation} 
where $1\leq\alpha,\beta \leq N$ and $\lambda_n$,  $\psi_{n}$ are the eigenvalues and the eigenvectors of the  Laplacian $L$ associated with the resistor network having unit resistance, that is, graphs. The  Laplacian matrix $L$  associated with the graph  $G=(V,E)$ is  $ L=D-A$, where  $D$ is the diagonal matrix of degrees   and $ A$ is the adjacency matrix  representing the edge set $E$. Now, let us give a suitable  method for  obtaining   the eigenvalues  for certain circulant graphs and in particular The eigenvalues of the  graph $ K_{N}^{-N}$. The complete graph minus $N$ edges is a  circulant graph \cite{Biggs}, that is,  a graph  whose vertices can be ordered so that the adjacency matrix $A$ is a circulant matrix, mathematically, this means that the $i^{\text{ith}}$ row is a cyclic shift of the $\text{zeroth}$ row by  $i$, $ a_{i,j} = a_{0,j-i}$, $ i,j=0,1,\cdots,N-1$, so that the vertices  $N$ and $0$ are identical, here, note that the subscripts are reduced modulo $N$. The graph  $ K_{N}^{-N}$ is an undirected graph  with no self-loops, i.e., $a_{i,j} = a_{ji}$ and $ a_{ii}=0$ respectively. Our graph belongs to the  the  undirected circulant graphs  denoted by $C_N(1,2,\cdots,p)$, where $p<N/2$, if $N$ is even, and $p<\frac{N-1}{2}$, if $N$ is odd, here, each vertex $i$ is  adjacent to $2p$  vertices  $ i \pm1 , i \pm2, \cdots, i \pm p$ mod $N$. In particular, if $p=1$, this is the $N$-cycle graph in which  each vertex is adjacent to two vertces, and if $p=N/2$, $p=\frac{N-1}{2}$, then, one obtains a complete graph whose number of vertices $N$ is even, odd respectively. Therefore, if $N$ is odd, the complete graph minus $N$ edges as a circulant graph is   $C_N(1,2,\cdots,\frac{N-1}{2}-1)$, so that each vertex is adjacent to $N-3$ vertices. The  matrix elements of the  Laplacian matrix $L$ of the graph $G$, may be written as   
\begin{eqnarray}
L_{mn}= \left\{\begin{array}{rl}
d & \text{if } m=n,\\
-1 & \text{if } \text {if m and n are adjacents} \\
0  & \text{otherwise}, 
\end{array} \right.
\end{eqnarray}
where $d$, is the degree of the graph that is, the number of unit resistors connected to  $i^{\text{th}}$ vertex. In order to obtain the eigenvalues of the Laplacian of the graph, it is more convenient to read the matrix elements of the adjacency matrix $A$ using circulant graphs. For example, the Laplacian for the $N$-cycle graph, the $N$-cycle with four nearest neighbors, and the complete graph minus $N$ edges for $N$ odd, may be written as 
 \begin{equation}
\label{a2}
L_{mn}=2\delta_{m,n}-(\delta_{m,n+1}+\delta_{m,n-1}),
\end{equation}
\begin{equation}
\label{a3}
L_{mn}=4\delta_{m,n}-(\delta_{m,n+1}+\delta_{m,n-1})-(\delta_{m,n+2}+\delta_{m,n-2}),
\end{equation}
and
 \begin{equation}
\label{a4}
L_{mn}=(N-3)\delta_{m,n}-(\delta_{m,n+1}+\delta_{m,n-1})-\cdots-(\delta_{m,n+(\frac{N-1}{2}-1)}+\delta_{m,n-(\frac{N-1}{2}-1)}),
\end{equation}
respectively. We should note that row and column labels of $\delta_{m,n\pm k} $ are to be considered modulo $N$. Since the matrix $ \Psi$ with elements $\psi_{n,k}=\frac{1}{\sqrt{N}}\exp(\frac{2\pi ink}{N}) $ is  a unitary transformation, this is a consequence of the identity $(\Psi^{*}\Psi)_{mn}=\frac{1}{N}\sum_{k=1}^{N}\exp[-2\pi ik(m-n)/N]=\delta_{m,n},$ then, the hermitian Laplacian matrix $L$  may be diagonalized using the matrix  $\Psi$. The matrix elements $\delta_{m,n+ k} $ may be considered as matrix elements of the   $k^{\text{th}}$ power of the rotation matrix $R$ of finite  closed lattice \cite{Wolf}, i.e, $(R^{k})_{mn}=\delta_{m,n+ k}$. As a consequence using $\Psi$,  $\delta_{m,n+ k} $ may be written as
\begin{eqnarray}
\label{a5}
(\mathcal R^{k})_{mn}&=&(\Psi^{*}R^{k}\Psi)_{mn}=\frac{1}{N}\sum_{j,l} \delta_{j,l+ k} \exp[2\pi i(ln-jm)/N] \nonumber\\&=&\exp[-2\pi ikm/N]\frac{1}{N}\sum_{l}\exp[2\pi i(n-m)/N]\nonumber\\&=& \delta_{m,n}\exp[-2\pi ikm/N],
\end{eqnarray}
similarly, $ (\mathcal R^{-k})_{mn}=\delta_{m,n}\exp[2\pi ikm/N]$. Now, this representation may be used to obtain the eigenvalues of the laplacian of any $N$-cycle graph with $2p$ nearest neighbors, for example the eigenvalues  for the $N$-cycle graph, the $N$-cycle graph with four nearest neighbors are $\lambda_{n}=4\sin^2(n\pi/N)$,   $ \lambda_n=4\sin^2(n\pi/N)+4\sin^2(2n\pi/N)$ respectively. In general, the eigenvalues of the $N$-cycle with $2p$ nearest neighbors are  $$ \lambda_n=4\sum_{m=1}^{p}\sin^2(nm\pi/N).$$ Thus, the eigenvalues of the Laplacian of the  complete graph minus $N$ edges are

\begin{equation}
\label{c1}
\lambda_{n}=4\sum_{m=1}^{\frac{N-1}{2}-1} \sin^2(nm\pi/N),
\end{equation}
by using the identity
\begin{equation}
\label{c2}
4\sum_{m=1}^{\frac{N-1}{2}} \sin^2(nm\pi/N)=N,
\end{equation}
it follows that the even and the odd eigenvalues are $$\lambda_{2n}=N-4 \sin^2(n\pi/N)$$ and $$\lambda_{2n-1}=N-4 \cos^2((2n-1)\pi/2N)$$
respectively. Substituting the expressions for the eigenvalues  $ \lambda_{n}$ of the Lapacian in Eq. (\ref{7}),  then the two point resistance of the   graph $ K_{N}^{-N}$ may be written as 
\begin{eqnarray} 
\label{c3}
R_{\alpha\beta}&=&R(|\alpha-\beta|)=R(l)\nonumber \\ &=&\frac{1}{N}\sum_{n=1}^{N-1}\frac{4\sin^2(nl\pi/N)}{N-4\sin^2(\frac{N-1}{2}n\pi/N)}\nonumber \\ &=&\frac{4}{N^2}\sum_{n=1}^{\frac{N-1}{2}}\frac{\sin^2(2nl\pi/N)}{1-\frac{4}{N}\sin^2(n\pi/N)}+\frac{4}{N^2}\sum_{n=1}^{\frac{N-1}{2}}\frac{\sin^2((2n-1)l\pi/N)}{1-\frac{4}{N}\cos^2((2n-1)\pi/2N)}.
\end{eqnarray}
 Therefore, to compute the two-point resistance we need to compute the last two sum that we denote by $R_{1}(l) $ and $R_{2}(l)$ respectively. The first term may be expanded to give
\begin{eqnarray} 
\label{c4}
R_{1}(l)&=&\frac{4}{N^2}\sum_{n=1}^{\frac{N-1}{2}}\frac{\sin^2(2nl\pi/N)}{1-\frac{4}{N}\sin^2(n\pi/N)}\nonumber\\ &=&\frac{2}{N^2}\sum_{j=0}^{\infty}(4/N)^{j}\Bigg(\sin^{2j}(n\pi/2N)\sum_{s=1}^{l}(-1)^{s+1} \frac{l}{l+s}  \binom {l+s} {l-s}2^{2s}\sum_{n=1}^{\frac{N-1}{2}}\sin^{2s}(2n\pi/N)\Bigg)\nonumber\\&=&\frac{1}{N^2}\sum_{j=0}^{\infty}(4/N)^{j}\Bigg(\sum_{s=1}^{l}(-1)^{s+1} \frac{l}{l+s}  \binom {l+s}{l-s}2^{4s}\sum_{m=0}^{s}(-1)^{m}\binom{s}{m}\sum_{n=1}^{N-1}\sin^{2(s+j+m)}(n\pi/N)\Bigg).\nonumber\\
\end{eqnarray}
 In obtaining the last two lines of the above equation we have used the trigonometrical identity $$\cos2(2ln\pi/N)=\sum_{s=0}^{l}(-1)^s \frac{l}{l+s}\binom {l+s} {l-s}2^{2s}\sin^{2s}(2n\pi/N), $$  and the symmetry $ \sin^{2J}n\pi/N=\sin^{2J}(N-1)n\pi/N$ for $ 1\leq n\leq N-1$. As it was explained in our recent paper \cite{Chair1}, that the suitable formula for the sum over $n$ is a slight deformation of the formula given by Schwatt \cite{Schwatt}. The suitable formula that we need is
\begin{eqnarray}
\label{c5}
\sum_{n=1}^{N-1}\sin^{2(s+j+m)}(n\pi/N)&=&
\frac{N}{2^{2(s+j+m)}}\binom {2(j+s+m)}{j+s+m}\nonumber\\&+&\frac{N}{2^{2(j+s+m)-1}}\sum_{p=1}^{[(j+s+m)/N]}(-1)^{p}\binom {2(j+s+m)}{j-pN}.
\end{eqnarray}
Therefore, the computation of  $R_{1}(l)$ splits into two parts  $R_{1}(l)^{'}$ and $R_{1}(l)^{''}$, where
\begin{eqnarray}
\label{c6}
R_{1}(l)^{'}=\frac{1}{N}\sum_{s=1}^{l}(-1)^{s+1} \frac{l}{l+s}  \binom {l+s}{l-s}(4N)^{s}\sum_{m=0}^{s}(-1)^{m}\binom{s}{m}(N/4)^m&\nonumber\\\times\Bigg(\sum_{J=0}^{\infty}(1/N)^{J}\binom {2J} {J}-\sum_{J=0}^{s+m-1}(1/N)^{J}\binom {2J} {J}\Bigg)
\end{eqnarray}
and
\begin{eqnarray}
\label{c7}
R_{1}(l)^{''}=\frac{1}{N}\sum_{s=1}^{l}(-1)^{s+1} \frac{2l}{l+s}  \binom {l+s}{l-s}(4N)^{s}\sum_{m=0}^{s}(-1)^{m}\binom{s}{m}(N/4)^m\nonumber\\\times\Bigg(\sum_{J=0}^{\infty}(1/N)^{J}\sum_{p=1}^{[J/N]}(-1)^{p}\binom {2J} {J-pN}-\sum_{J=0}^{s+m-1}(1/N)^{J}\sum_{p=1}^{[J/N]}(-1)^{p}\binom {2J} {J-pN}\Bigg),
\end{eqnarray}
where  $J=j+s$.
The next thing to do is to evaluate  the sum over $J$, to do so we use the representation of the  binomial coefficients by residue \cite{Egorychev}. First, let us recall the definition of the residue operator, if $G(w)= \sum_{k=0}^{\infty}a_{k}w^{k}$ is a generating function for a sequence $\{a_{k}\}$. Then the k-th coefficient of $G(w)$ may be represented by the formal residue as follows
$$a_{k}= \hbox{res}_w G(w){w^{-k-1}}.$$
In particular, the generating function of the binomial coefficient sequence  $\binom {n}{k}$ for a fixed $n$ is given by $$G(w)= \sum_{k=0}^{n}\binom {n}{k} w^{k} =(1+w)^{n},$$ 
and hence $$\binom {n}{k}=\hbox{res}_w (1+w){^n}{w^{-k-1}}.$$
The other binomial coefficient that we need is the one that in which $n$ takes all integer values, like the first sum over $J$ given in Eq. (\ref{c6}), this particular binomial coefficient is given by $$\binom {2n}{n}=\hbox{res}_w (1-4w){^{-1/2}}{w^{-n-1}}.
$$ 
Before finishing this brief summary, we should mention  one important property of the  residue operator $\hbox{res}$, namely linearity. This is crucial in our computations,  linearity states that  given some contants $\alpha$ and $\beta$, then $$\alpha \hbox{res}_w G_{1}(w){w^{-k-1}}+ \beta\hbox{res}_w G_{2}(w){w^{-k-1}}=\hbox{res}_w(\alpha G_{1}(w)+ \beta G_{2}(w)) {w^{-k-1}}.$$ Therefore, the first sum over $J$ in Eq. (\ref{c6}) may be written as
 \begin{equation}
\label{c8}
\sum_{J=0}^{\infty}(1/N)^{J}\binom {2J} {J}=\hbox{res}_w (1-4w){^{-1/2}}\sum_{J=0}^{\infty}(1/Nw)^{J}{w^{-1}}=\frac{\sqrt{N}}{\sqrt{N-4}}
\end{equation}
 As a result the first term of Eq. (\ref{c6}) may be written as 
 \begin{eqnarray}
 \label{c9}
\frac{1}{N}\sum_{s=1}^{l}(-1)^{s+1} \frac{l}{l+s}  \binom {l+s}{l-s}(4N)^{s}\sum_{m=0}^{s}(-1)^{m}\binom{s}{m}(N/4)^m\sum_{J=0}^{\infty}(1/N)^{J}\binom {2J} {J}&\nonumber\\=\frac{1}{2N}\frac{\sqrt{N}}{\sqrt{N-4}}\sum_{k=0}^{l-1}(-1)^{l+k+1}\frac{2l}{2l-k}\binom{2l-k}{k}\bigg(\sqrt{N(4-N)}\bigg)^{2(l-k)},
 \end{eqnarray}
 where we have set $ l-s=k$.  Now, the  sum over $k$ of Eq. (\ref{c9})  is nothing but the normalized  Chebyshev polynomial of the first kind $C_{2l}\Big(\sqrt{N(4-N)}\Big) $ without the term $k=l$, where $$ C_{2l}(x)= 2T_{2l}(x/2)=\sum_{k=0}^{l-1}(-1)^{k} \frac{2l}{2l-k}\binom {2l-k} {k}x^{2l-2k},$$
is the normalized Chebyshev polynomial \cite{Rivlin}, and
$$T_{2l}(x/2)=\frac{1}{2}\Bigg\lbrack\Bigg(\frac{x}{2}+\sqrt{(x/2)^2-1}\bigg)^{2l} +\Bigg(\frac{x}{2}-\sqrt{(x/2)^2-1}\bigg)^{2l} \Bigg\rbrack.$$
Therefore, equation (\ref{c9}) may be written as 
\begin{eqnarray}
\label{c10}
\frac{1}{2N}\frac{\sqrt{N}}{\sqrt{N-4}}\sum_{k=0}^{l-1}(-1)^{l+k+1}\frac{2l}{2l-k}\binom{2l-k}{k}\bigg(\sqrt{N(4-N)}\bigg)^{2(l-k)}\nonumber\\=\frac{1}{2N}\frac{\sqrt{N}}{\sqrt{N-4}}(-1)^{l+1}\Big(C_{2l}\Big(\sqrt{N(4-N)}-2(-1)^{l}\Big)\nonumber\\=-\frac{1}{2}\frac{1}{\sqrt{N(N-4)}}\Bigg(\Big(\frac{N-2+\sqrt{N(N-4)}}{2}\Big)^{2l}+\Big(\frac{N-2-\sqrt{N(N-4)}}{2}\Big)^{2l}-2\Bigg)\nonumber\\=-\frac{1}{2}\sqrt{N(N-4)}\mathcal{B}_{l}^{2}(N),
\end{eqnarray}
where $ \mathcal{B}_{l}(N)$ are Bejaia numbers that are defined below. In order to make our formulas nicer and less complicated,  we  introduce certain numbers $  \mathcal{B}_{l}(N)$, that we call the Bejaia numbers   given by  the following expression
\begin{equation}
\label{B}
 \mathcal{B}_{l}(N)=\frac{1}{\sqrt{N(N-4)}}\Bigg(\Big(\frac{N-2+\sqrt{N(N-4)}}{2}\Big)^{l}-\Big(\frac{N-2-\sqrt{N(N-4)}}{2}\Big)^{l}\Bigg).
\end{equation}
Then,  $$ \mathcal{B}_{l}^{2}(N)= \frac{1}{{N(N-4)}}\Bigg(\Big(\frac{N-2+\sqrt{N(N-4)}}{2}\Big)^{2l}+\Big(\frac{N-2-\sqrt{N(N-4)}}{2}\Big)^{2l}-2\Bigg),$$ this is  reminiscent of the relation $ F_{l}^{2}=\frac{{1}}{5}\bigg(L_{2l}-2(-1)^{l} \bigg)$, where $F_{l}$ and $ L_{2l}$ are the Fibonacci and the Lucas numbers respectively.
Therefore, we may define the Bejaia's cousin numbers $ \mathcal{P}_{l}(N)$, that we call the Pisa numbers defined  by
 \begin{equation}
\label{P}
\mathcal{P}_{l}(N)= \Big(\frac{N-2+\sqrt{N(N-4)}}{2}\Big)^{l}+\Big(\frac{N-2-\sqrt{N(N-4)}}{2}\Big)^{l},
\end{equation} 
so that $\mathcal{B}_{l}^{2}(N)= \frac{1}{{N(N-4)}}(\mathcal{P}_{2l}(N)-2)$. The Bejaia numbers,   and the Pisa numbers  are  generalizations  of the bisected Fibonacci and the Lucas numbers respectively. For $ N=5$, then the Bejaia numbers $ \mathcal{B}_{l}(5)$ are nothing but the bisection of the  Fibonacci numbers, $F_{2l}$, and similarly, the Pisa numbers $ \mathcal{P}_{l}(5)$,  are the bisection of the Lucas numbers, that is,   $ L_{2l}$, see sequences ($ A001906 $, $A005248 $) \cite{Sloane}. 
We now go to the second term of Eq. (\ref{c6}), this time the summation over $J$ is a finite sum and the suitable residue representation is
\begin{eqnarray}
\label{c11}
\sum_{J=0}^{s+m-1}(1/N)^{J}\binom {2J} {J}=\sum_{J=0}^{s+m-1}(1/N)^{J}\hbox{res}_{w=0} (1+w){^{2J}}{w^{-J-1}}\nonumber\\=N\hbox{res}_{w=0}\frac{(1+w)^{2(s+m)}}{(Nw)^{s+m}((1+w)^{2}-Nw)}
\end{eqnarray}
Using the above equation and summing over $m$, and then evaluating the residue, the second term of Eq.  (\ref{c6}) becomes
\begin{eqnarray}
\label{c12}
&&\frac{1}{N}\sum_{s=1}^{l}(-1)^{s} \frac{l}{l+s}\binom {l+s}{l-s}(4N)^{s}\sum_{m=0}^{s}(-1)^{m}\binom{s}{m}(N/4)^m\sum_{J=0}^{s+m-1}(1/N)^{J}\binom {2J}{J}\nonumber\\&=&\frac{1}{2}\hbox{res}_{w=0}\frac{1}{(w-(\frac{N-4)-\sqrt{N(N-4)}}{2})(w-(\frac{N-4)+\sqrt{N(N-4)}}{2})}\Big(C_{2l}\Big(\frac{i(1-w^2)}{w}\Big)-2(-1)^{l}\Big)\nonumber\\&=&\frac{\mathcal{B}_{2l}(N)}{2}
\end{eqnarray}
In order to compute the first term of Eq. (\ref{c7}), we first,  sum over $J$ which is  done through residue representation  as follows
\begin{eqnarray}
\label{c13}
\sum_{J=0}^{\infty}(1/N)^{J}\sum_{p=1}^{[J/N]}(-1)^{p}\binom  {2J} {J-pN}&=&-\hbox{res}_{w}\sum_{J=0}^{\infty}(1/N)^{J}(1+w)^{2J}w^{-J-1}\Big(\frac{w^{N}}{1+w^{N}} \Big)\nonumber\\&=&-\frac{N}{\sqrt{N(N-4)}}\frac{\Big(\frac{N-2-\sqrt{N(N-4)}}{2}\Big)^{N}}{1+\Big(\frac{N-2-\sqrt{N(N-4)}}{2}\Big)^{N}},
\end{eqnarray}
in obtaining Eq. ( \ref{c13}), we allowed the sum over $p$ to go to infinity, since for $p >[J/N]$, the binomial is identically zero and subtracted the $p=0$ term from the infinite geometrical series $ \sum_{p=0}^{\infty}(-1)^{p}(w^{N})^{p} $, $ |{w}|<1$.
Then,  the sum of the first term in  Eq. (\ref{c7}) reads
\begin{eqnarray}
\label{c14}
\frac{1}{N}\sum_{s=1}^{l}(-1)^{s+1} \frac{2l}{l+s} \binom {l+s}{l-s}(4N)^{s}\sum_{m=0}^{s}(-1)^{m}\binom{s}{m}(N/4)^m\sum_{J=0}^{\infty}(1/N)^{J}\sum_{p=1}^{[J/N]}(-1)^{p}\binom {2J} {J-pN}\nonumber\\=\sqrt{N(N-4)}\frac{\Big(\frac{N-2-\sqrt{N(N-4)}}{2}\Big)^{N}}{1+\Big(\frac{N-2-\sqrt{N(N-4)}}{2}\Big)^{N}}\mathcal{B}_{l}^{2}(N).
\end{eqnarray}
Finally, we come to the last computation of the first part of the the two-point resistance, that is, the second term in Eq. (\ref{c7}). This time the sum over $J$ is finite and one has
\begin{eqnarray}
\label{c15}
\sum_{J=0}^{s+m-1}(1/N)^{J}\sum_{p=1}^{[J/N]}(-1)^{p}\binom  {2J} {J-pN}=\frac{w^{N}}{1+w^{N}}\hbox{res}_{w=0}\frac{(1+w)^{2(s+m)}}{(Nw-(1+w^2))(Nw)^{s+m}}.
\end{eqnarray}
In the above equation we have discarded a term whose residue at $ w=N-2-\sqrt{N^{2}-4N}$ do not contribute and has a vanishing residue at $ w=0$. Summing over $m$, then,  the residue computation at the pole $w=0$ of order $ 2l-N$ gives
\begin{eqnarray}
\label{c16}
&&\frac{1}{N}\sum_{s=1}^{l}(-1)^{s} \frac{2l}{l+s}  \binom {l+s}{l-s}(4N)^{s}\sum_{m=0}^{s}(-1)^{m}\binom{s}{m}(N/4)^m\sum_{J=0}^{s+m-1}(1/N)^{J}\sum_{p=1}^{[J/N]}(-1)^{p}\binom {2J} {J-pN}\nonumber\\&=&-\mathcal{B}_{2l-N}(N).
\end{eqnarray}
It is interesting to note that this term  contributes only for $ l\ge (N+1)/2$. As a result,  now, adding the different contributions given by equations (\ref{c10}), (\ref{c12}), and (\ref{c16}), then, the first term $R_{1}(l) $ of the two-point resistance of the  graph $ K_{N}^{-N}$ has  the following closed formula
\begin{eqnarray}
\label{c17}
\label{R_{1}(l)}
R_{1}(l)=\frac{4}{N}\sum_{n=1}^{\frac{N-1}{2}}\frac{\sin^2(2nl\pi/N)}{N-4\sin^2(n\pi/N)}=\frac{\mathcal{B}_{2l}(N)}{2}-\frac{\sqrt{N(N-4)}}{2}\mathcal{B}_{l}^{2}(N)\frac{1-\Big(\frac{(N-2-\sqrt{N(N-4)}}{2}\Big)^{N}}{1+\Big(\frac{N-2-\sqrt{N(N-4)}}{2}\Big)^{N}}
\end{eqnarray}
The computations of the second term of the two-point resistance  $ R(l)$,  namely, $ R_{2}(l)$, are to a certain extent similar to those of $ R_{1}(l)$, the details of these computations  will be given in appendix \ref{Half}  . Although, not all the contributions are the same as in $ R_{1}(l)$, the total contribution give exactly the same results obtained for  $ R_{1}(l)$, then, the second term $ R_{2}(l)$ is
\begin{eqnarray}
\label{c181}
R_{2}(l)&=&\frac{4}{N}\sum_{n=1}^{\frac{N-1}{2}}\frac{\sin^2((2n-1)l\pi/N)}{N-4\cos^2((2n-1)\pi/2N)}\nonumber\\&=&\frac{\mathcal{B}_{2l}(N)}{2}-\frac{\sqrt{N(N-4)}}{2}\mathcal{B}_{l}^{2}(N)\frac{1-\Big(\frac{(N-2-\sqrt{N(N-4)}}{2}\Big)^{N}}{1+\Big(\frac{N-2-\sqrt{N(N-4)}}{2}\Big)^{N}}.
\end{eqnarray}
Therefore, the final expression for the two-point resistance of the graph $ K_{N}^{-N}$, reads;
\begin{eqnarray}
\label{c26}
R(l)&=&\frac{1}{N}\sum_{n=1}^{N-1}\frac{4\sin^2(nl\pi/N)}{N-4\sin^2(\frac{N-1}{2}n\pi/N)} =\nonumber\\& =&\mathcal{B}_{2l}(N)-\sqrt{N(N-4)}\mathcal{B}_{l}^{2}(N)\frac{1-\Big(\frac{(N-2-\sqrt{N(N-4)}}{2}\Big)^{N}}{1+\Big(\frac{N-2-\sqrt{N(N-4)}}{2}\Big)^{N}},
\end{eqnarray}
Where $N>4$, and $l<\frac{N+1}{2}$.
\section{The total effective resistance, and the simple random walk on the complete graph minus $N$  edges }
Effective resistances in  electrical networks, have been known to have far reaching implications in a variety of problems. Recurrence and transience in random walks
in infinite networks Doyle and  Snell \cite{Doyle},
and the covering and commute times of random walks in graphs \cite{Tetali, Chandra} can be determined by the
effective resistance. A simple random walk, on a graph  is  conveniently represented by its transition  probability $ P_{ij}=\frac{1}{d(i)}$, if $i$, $j$ are adjacent vertices and $0$ otherwise (where $d(i)$ is the degree of $i$). This is the probability the walk  move from vertex $i$ to  vertex $j$, given that we are at vertex $i$. There are important parameters associated with random walks on graphs  such as the (FPT) and  and commute time that  may be written in terms of the effective resistance. The (FPT) $ H_{ij}$ is the expected number of steps it takes  a walk that starts at $i$ to reach $j$. The commute time $C_{ij}$ is the expected number of steps that it takes  a walk to go from $i$ to  $j$ and back to $ i$ so $ C_{ij}=H_{ij}+H_{ji}$.   It has been proved by Chandra et al. \cite{ Chandra} that the commute time $ C_{ij }$ is equal to $2 \vert E\vert R_{ij}$, and from the total resistance distance of the graph $ G$, introduced by Klein and Randic \cite{randic} $  R(G)= \sum_{i<j}R_{ij}$, then 
 \begin{equation}
\label{com1} R(G)=\frac{1}{2\vert E\vert}\sum_{i<j}(H_{ij}+H_{ji})=\frac{1}{2\vert E\vert}\sum_{ij}H_{ij}
\end{equation}
that, is, the total effective resistance of a graph is proportional to the expected commute time averaged over all pairs of vertices. Given a closed form  expression for the  effective resistance, then  the total effective resistance $ R(G$, the (FPT), and (MFPT), may be obtained in closed forms. The expression for (FPT)  of the simple random walk on he complete graph minus $N$  edges, may be written as $H_{0,l}=\vert E\vert R(l)$, since the graph is regular and has a rotational symmetry. Now, the total number of edges in this graph is $\frac{N(N-3)}{2}$, therefore, the exact expression for the (FPT) of the simple random walk on the graph $K_{N}^{-N}, $ reads
\begin{equation}
\label{hit1}
H_{0,l}=\frac{N(N-3)}{2}\Bigg(\mathcal{B}_{2l}(N)-\sqrt{N(N-4)}\mathcal{B}_{l}^{2}(N)\frac{1-\Big(\frac{(N-2-\sqrt{N(N-4)}}{2}\Big)^{N}}{1+\Big(\frac{N-2-\sqrt{N(N-4)}}{2}\Big)^{N}}\Bigg).
\end{equation}
In order to compute the (MFPT), we need  an expression for the total effective resistance of this  graph, this may be seen as follows, 
 \begin{equation} 
 \label{mean first hit}
 \overline{H_{0,l}}=\frac{1}{N}\sum_{l=1}^{N-1} H_{0,l} =\frac{2\vert E\vert}{N^2}\sum_{l=1}^{N-1}\frac{N}{2}R(l)=\frac{d}{N}R(G),
\end{equation}
 where we have used the formula $2\vert E\vert= Nd$ for regular graphs, here, the degree of the graph is $d=N-1$. The expression of the total effective resistance may be obtained  in a closed form  by using the expressions for the sums $ \sum_{l=1}^{N-1}\mathcal{B}_{2l}(N)$ and $ \sum_{l=1}^{N-1}\mathcal{B}_{l}^{2}(N)$ evaluated in Appendix\ref{sumsB}. As a consequence, the total resistance may be computed to give
\begin{eqnarray}
\label{total resistance}
R(K_{N}^{-N})&=&N\sum_{l=1}^{\frac{N-1}{2}}R(l)\nonumber\\&=&N\Bigg(\frac{\mathcal{P}_{N}(N)-(N-2)}{N(N-4)}-\frac{\mathcal{B}_{N}(N)-N}{\sqrt{N(N-4)}}\frac{1-\Big(\frac{(N-2-\sqrt{N(N-4)}}{2}\Big)^{N}}{1+\Big(\frac{N-2-\sqrt{N(N-4)}}{2}\Big)^{N}}\Bigg),
\end{eqnarray}
in obtaining the above formula for the total resistance we used the fact that $\mathcal{B}_{1}(N)=1 $, and   $\mathcal{P}_{1}(N)=N-2$. Thus, the (MFPT) of the simple random walk on $K_{N}^{-N} $, is
\begin{equation}
\label{mean hitting time}
\overline{H_{0,l}}=(N-1)\Bigg(\frac{\mathcal{P}_{N}(N)-(N-2)}{N(N-4)}-\frac{\mathcal{B}_{N}(N)-N}{\sqrt{N(N-4)}}\frac{1-\Big(\frac{(N-2-\sqrt{N(N-4)}}{2}\Big)^{N}}{1+\Big(\frac{N-2-\sqrt{N(N-4)}}{2}\Big)^{N}}\Bigg).
\end{equation}
It has been shown in \cite{randic}, that the total effective resistance of a connected graph $G$  with $N$ vertices may be written in terms of the Laplacian eigenvalues as $ R(G)=N\sum_{n=1}^{N-1}\frac{1}{\lambda_n}$,  therefore, the equivalent formula for the (MFPT) of the simple random walk on the graph $ K_{N}^{-N}$, is
\begin{equation}
\label{mean hit equi}
\overline{H_{0,l}}=(N-1)\sum_{n=1}^{N-1}{\Bigl{(}4\sum_{m=1}^{\frac{N-1}{2}-1}\sin^2mn\pi/N}\Bigr{)}^{-1}
\end{equation}
Comparison with eq. (\ref{mean hitting time}) entitles us to draw the conclusion that we should have the following identity
\begin{eqnarray}
\label{eigentime identity}
\sum_{n=1}^{N-1}{\Bigl{(}4\sum_{m=1}^{\frac{N-1}{2}-1}\sin^2mn\pi/N}\Bigr{)}^{-1}&=&\Bigg(\frac{\mathcal{P}_{N}(N)-(N-2)}{N(N-4)}-\frac{\mathcal{B}_{N}(N)-N}{\sqrt{N(N-4)}}\frac{1-\Big(\frac{(N-2-\sqrt{N(N-4)}}{2}\Big)^{N}}{1+\Big(\frac{N-2-\sqrt{N(N-4)}}{2}\Big)^{N}}\Bigg).\nonumber\\
\end{eqnarray}
For example, we may check the above identity for the graph   $K_{7}^{-7}$, i.e., the  $7-cycle$ graph in which every vertex is adjacent to $4$ nearest neighbors. In this case the   Bejaia and the Pisa numbers  are $\mathcal{B}_{7}(7)= 12649 $, and  $\mathcal{P}_{7}(7)=57965$ respectively, see Appendix  \ref{BP}, then, computing  both sides  of Eq. (\ref{eigentime identity}) give the value $ 1.3846153$.  
\section{Conclusion}
To conclude, in this work, we were able to obtain the exact formula for the two-point resistance, the total effective resistance, the (FPT)  and  the (MFPT) of the simple random walk on the the complete graph minus  $ N$  edges of the opposite vertices $K_{N}^{-N}$. These  formulas are written in terms of  certain numbers that we called the Bejaia, and the Pisa numbers that are generalizations of the bisected Fibonacci and the Lucas numbers. By uncovering the properties of the Bejaia, and the Pisa numbers, then the exact two-point resistance may be obtained for any number of vertices $N$ of the graph  $K_{N}^{-N}$ . These numbers were shown to be related to each other through some identities, similar to the relations between the Fibonacci and the Lucas numbers. Using these identities, then the sum of the Bejaia and the Pisa numbers and their powers were possible and given in closed forms  which played a crucial role in obtaining the exact formulas for the total effective resistance, the (FPT),  and  the (MFPT) of the simple random walk on the the graph   $K_{N}^{-N}$. The connection between the random walk and the bisected Fibonacci numbers have been already noticed in the literatures \cite{Stein}, If, one consider a particle  executing random walk on the line that starts at the point $1$ and arrives eventually at the point $5$ in a total of $4+2l$ probable unit steps, $l$  of which are in the negative direction. Then,  the number  of distinct walks satisfying these restrictions is given by the bisection of the Fibonacci numbers (alternate Fibonacci numbers) . The same random walk is generated by computing the two-point correlators (intersection numbers) on the moduli space of holomorphic maps, of a fixed degree $d$, from a  sphere into the Grassmannian of $ 2$-planes  in $\mathbb{C}^5 $\cite{Chair}. In a recent work \cite{Chair1},  the exact two-point resistance  of  the exact  expression for  the two-point resistance  of the square of the $N$-cycle graph, $ C_{N}(1,2)$, in which every vertex is connected to its two neighbors and neighbor's neighbors, in which it was shown to be written in terms of  two-point resistance of the $N$-cycle graph $C_{N}$,  the square of the Fibonacci numbers, and the bisected Fibonacci numbers. Therefore, the two-point resistance  of the graph $K_{N}^{-N}$, in this paper generalizes naturally  the two-point resistance of the graph $ C_{N}(1,2)$. In general, the important parameters of  random walks on circulant graphs are related to the Fibonacci numbers and their generalizations.
\appendix
\section{The explicit computation of the sum  $R_{2}(l)$}\label {Half}
Here, we will show that the sun $R_{2}(l)$ given in  Eq. (\ref{c18}) is exactly equal to $R_{1}(l)$, see Eq. (\ref{c17}. The expression for  $R_{2}(l)$ may be written as
\begin{eqnarray}
\label{c18}
R_{2}(l)&=&\frac{4}{N}\sum_{n=1}^{\frac{N-1}{2}}\frac{\sin^2((2n-1)l\pi/N)}{N-4\cos^2((2n-1)\pi/2N)}\nonumber\\&=&\frac{2}{N^2}\sum_{j=0}^{\infty}(4/N)^{j}\sum_{s=1}^{l}(-1)^{s+1} \frac{l}{l+s}  \binom {l+s}{l-s}2^{4s}\nonumber\\&\times&\Bigg(\sum_{m=0}^{s}(-1)^{m}\binom{s}{m}\sum_{n=1}^{\frac{N-1}{2}}\cos^{2(s+j+m)}((2n-1)\pi/2N)\Bigg).
\end{eqnarray}
Using the following identity
\begin{eqnarray}
\label{psum}
\sum_{n=1}^{\frac{N-1}{2}}\cos^{2(s+j+m)}((2n-1)\pi/2N)&=&\frac{1}{2}\sum_{n=1}^{N-1}\cos^{2(s+j+m)}(n\pi/2N)\nonumber\\&+&\frac{1}{2}\sum_{n=1}^{N-1}(-1)^{n-1}\cos^{2(s+j+m)}(n\pi/2N),
\end{eqnarray}
one can show that the  suitable formula for the power sums in  Eq. (\ref{psum}), that take into account that $j$ may be congruent to $N$, is 
\begin{eqnarray}
\label{c19}
\sum_{n=1}^{\frac{N-1}{2}}\cos^{2(s+j+m)}((2n-1)l\pi/N))&=&
\frac{N}{2^{2(s+j+m)+1}}\binom {2(j+s+m)}{j+s+m}\nonumber\\&+&\frac{N}{2^{2(j+s+m)}}\sum_{p=1}^{[(j+s+m)/2N]}\binom {2(j+s+m)}{j-2pN}\nonumber\\&-&\frac{N}{2^{2(j+s+m)}}\sum_{p=1}^{[(j+s+m)/2N]}\binom {2(j+s+m)}{j-(2p-1)N}.
\end{eqnarray}
The terms on the right hand of Eq. (\ref{c19}) are similar to those on the right of Eq. (\ref{c5}), the only difference is that this time we have sums over even and odd $p$ without the alternating factor $(-1)^{p}$. Therefore, the corresponding sums over $J$ may be computed to give
\begin{eqnarray}
\label{c20}
\sum_{J=0}^{\infty}(1/N)^{J}\sum_{p=1}^{[J/2N]}\binom  {2J} {J-2pN}&=&\hbox{res}_{w}\sum_{J=0}^{\infty}(1/N)^{J}(1+w)^{2J}w^{-J-1}\Big(\frac{w^{2N}}{1-w^{2N}} \Big)\nonumber\\&=&\frac{N}{\sqrt{N(N-4)}}\frac{\Big(\frac{N-2-\sqrt{N(N-4)}}{2}\Big)^{2N}}{1-\Big(\frac{N-2-\sqrt{N(N-4)}}{2}\Big)^{2N}},
\end{eqnarray}
and 
\begin{eqnarray}
\label{c21}
\sum_{J=0}^{\infty}(1/N)^{J}\sum_{p=1}^{[J/2N]}\binom  {2J} {J-(2p-1)N}&=&\hbox{res}_{w}\sum_{J=0}^{\infty}(1/N)^{J}(1+w)^{2J}w^{-J-1}\Big(\frac{w^{N}}{1-w^{2N}} \Big)\nonumber\\&=&\frac{N}{\sqrt{N(N-4)}}\frac{\Big(\frac{N-2-\sqrt{N(N-4)}}{2}\Big)^{N}}{1-\Big(\frac{N-2-\sqrt{N(N-4)}}{2}\Big)^{2N}},
\end{eqnarray}
respectively. The first sum over $J$  in Eq. (\ref{c18}) is exactly $R_{1}(l)^{'}$, this was already computed, that is,  the sum of the two contributions given by  Eqs. (\ref{c10}) and (\ref{c12}) respectively. The second and the third sums  over $J$  may be computed using the previous computations and the last two equations, to obtain
\begin{eqnarray}
\label{c22}
\frac{1}{N}\sum_{s=1}^{l}(-1)^{s+1} \frac{2l}{l+s}  \binom {l+s}{l-s}(4N)^{s}\sum_{m=0}^{s}(-1)^{m}\binom{s}{m}(N/4)^m\sum_{J=0}^{\infty}(1/N)^{J}\sum_{p=1}^{[J/2N]}\binom {2J} {J-pN}&=&\nonumber\\-\sqrt{N(N-4)}\frac{\Big(\frac{N-2-\sqrt{N(N-4)}}{2}\Big)^{2N}}{1-\Big(\frac{N-2-\sqrt{N(N-4)}}{2}\Big)^{2N}}\mathcal{B}_{l}^{2}(N),
\end{eqnarray}
and
\begin{eqnarray}
\label{c23}
\frac{1}{N}\sum_{s=1}^{l}(-1)^{s} \frac{2l}{l+s}  \binom {l+s}{l-s}(4N)^{s}\sum_{m=0}^{s}(-1)^{m}\binom{s}{m}(N/4)^m\sum_{J=0}^{\infty}(1/N)^{J}\sum_{p=1}^{[J/2N]}\binom {2J} {J-(2p-1)N}&=&\nonumber\\\sqrt{N(N-4)}\frac{\Big(\frac{N-2-\sqrt{N(N-4)}}{2}\Big)^{N}}{1-\Big(\frac{N-2-\sqrt{N(N-4)}}{2}\Big)^{2N}}\mathcal{B}_{l}^{2}(N),
\end{eqnarray}
respectively. Note that in obtaining  the second and the third sums given by Eqs. (\ref{c22}), (\ref{c23}) respectively,  we used the fact that $$\sum_{J=0}^{s+m-1}(1/N)^{J}\sum_{p=1}^{[J/2N]}\binom  {2J} {J-2pN}=\hbox{res}_{w}\sum_{J=0}^{\infty}(1/N)^{J}(1+w)^{2J}w^{-J-1}\Big(\frac{w^{2N}}{1-w^{2N}} \Big), $$
does not contribute to the residue, whereas the sum
$$ \sum_{J=0}^{s+m-1}(1/N)^{J}\sum_{p=1}^{[J/2N]}\binom  {2J} {J-(2p-1)N}=\hbox{res}_{w}\sum_{J=0}^{\infty}(1/N)^{J}(1+w)^{2J}w^{-J-1}\Big(\frac{w^{N}}{1-w^{2N}} \Big),$$
does contribute to the residue and hence to the third sum provided $ l\ge (N+1)/2$. The result of this computation coincides with the one obtained earlier see Eq. (\ref{c16}), that is,
\begin{eqnarray}
\label{c24}
&&\frac{1}{N}\sum_{s=1}^{l}(-1)^{s} \frac{2l}{l+s} \binom {l+s}{l-s}(4N)^{s}\sum_{m=0}^{s}(-1)^{m}\binom{s}{m}(N/4)^{m}\nonumber\\&\times&\sum_{J=0}^{s+m-1}(1/N)^{J}\sum_{p=1}^{[J/2N]}\binom {2J} {J-(2p-1)N}=-\sqrt{N(N-4)}\mathcal{B}_{(2l-N)}(N).
\end{eqnarray}
Finally, adding all the contributions given  by equations (\ref{c10}), (\ref{c12}), (\ref{c22}),  and (\ref{c23}), then, the second term of the two-point resistance $ R_{2}(l)$ of the graph $K_{N}^{-N} $ may be written in a closed form as follows
\begin{eqnarray}
\label{c25}
R_{2}(l)&=&\frac{4}{N}\sum_{n=1}^{\frac{N-1}{2}}\frac{\sin^2((2n-1)l\pi/N)}{N-4\cos^2((2n-1)\pi/2N)}\nonumber\\& =&\frac{\mathcal{B}_{2l}(N)}{2}-\frac{\sqrt{N(N-4)}}{2}\mathcal{B}_{l}^{2}(N)\frac{1-\Big(\frac{(N-2-\sqrt{N(N-4)}}{2}\Big)^{N}}{1+\Big(\frac{N-2-\sqrt{N(N-4)}}{2}\Big)^{N}}.
\end{eqnarray}
Therefore, the sums  $ R_{1}(l)$, and $ R_{2}(l)$ are identical. 
\section{The general properties of the Bejaia and the Pisa numbers}\label{BP}
The Bejaia  numbers may be shown to satisfy the following  recursion,  $\mathcal{B}_{l}(N)=(N-2) \mathcal{B}_{l-1}(N)-\mathcal{ B}_{l-2}(N) $, $ l\geq2$, therefore, for any $N\geq5$, one has,  $ \mathcal{B}_{0}(N)=0$, $ \mathcal{B}_{1}(N)=1$,  $\mathcal{ B}_{2}(N)=N-2 $, $\mathcal {B}_{3}(N)=(N-2 )^{2}-1 $, etc. Similarly, the Pisa numbers satisfy  $\mathcal{P}_{l}(N)=(N-2) \mathcal{P}_{l-1}(N)-\mathcal{ P}_{l-2}(N) $, $ l\geq2$, then, for any $N\geq 5$, one has $ \mathcal{P}_{0}(N)=2$, $ \mathcal{P}_{1}(N)=N-2$, $\mathcal {P}_{2}(N)=(N-2 )^{2}-2 $, $\mathcal{ P}_{3}(N)=(N-2)^3 -3(N-2)$, etc. As a consequence, we have an infinity number of the Bejaia and the Pisa sequences and  the two-point resistance may be evaluated for any $ N$, and $l$. The left hand of the above  trigonometrical sum enjoys the transparent symmetry $ R(l)=R(N-l)$, whereas, on the right hand side this symmetry is not manifest. This symmetry, is checked explicitly by taking into account  Eq.  (\ref{c16}), for  $ l\ge (N+1)/2$. This in turn implies that the Bejaia numbers satisfies the following identity 
\begin{equation}
\label{c29}
\mathcal{B}_{2(N-l)}(N)-\mathcal{B}_{2l}(N)-2\mathcal{B}_{(N-2l)}(N)=\sqrt{N(N-4)}\frac{1-\Big(\frac{(N-2-\sqrt{N(N-4)}}{2}\Big)^{N}}{1+\Big(\frac{N-2-\sqrt{N(N-4)}}{2}\Big)^{N}}\big(\mathcal{B}_{N-l}^{2}(N)-\mathcal{B}_{l}^{2}(N)\big).
\end{equation}
This is checked to be correct using simply the definition of $ \mathcal{B}_{l}(N)$ given in Eq. (\ref{B}).  
 
\subsection{Closed formulas for $ \sum_{l=1}^{N-1}\mathcal{B}_{l}(N)$ and $ \sum_{l=1}^{N-1}\mathcal{B}_{l}^{2}(N)$}\label{sumsB}
In order to obtain  the expression of the total effective resistance  in a closed form, we need to evaluate the following sums $ \sum_{l=1}^{N-1}\mathcal{B}_{l}(N)$ and $ \sum_{l=1}^{N-1}\mathcal{B}_{l}^{2}(N)$. This turns out to be possible through certain identities satisfied by the Bejaia and the Pisa numbers, these identities are the analog of the Fibonacci and the Lucas numbers. From the expressions for the Bejaia and Pisa  numbers, one obtains the following identities;$$\mathcal{B}_{l}(N)=\frac{1}{N(N-4)}\big(\mathcal{P}_{l+1}(N)-\mathcal{P}_{l-1}(N) \big) ,$$ and  $$\mathcal{P}_{l}(N)=\mathcal{B}_{l+1}(N)-\mathcal{B}_{l-1}(N) ,$$ 
from these idntities, we find 
\begin{equation}
\label{power sum1}
\sum_{l=1}^{n}\mathcal{B}_{l}^{2}(N)=\frac{1}{N(N-4)}\big(\mathcal{B}_{2n+1}(N)-\mathcal{B}_{1}(N)-2n\big), 
 \end{equation}
 and
\begin{equation}
\label{cousin sum}
\sum_{l=1}^{n}\mathcal{B}_{2l}(N)=\frac{1}{N(N-4)}\big(\mathcal{P}_{2n+1}(N)-\mathcal{P}_{1}(N)\big). 
\end{equation}
\newpage
\bibliographystyle{phaip}

\begin{thebibliography}{1}
\bibitem{Doyle} P.G. Doyle, J.L. Snell, Random Walks and Electrical Networks, The Mathematical Association of America, Washington, DC, 1984
\bibitem{Tetali} P. Tetali,  (1991). Journal of Theoretical Probability, 4, 101 (1991)
\bibitem{Chandra} A.K Chandra, P. R  Raghavan, W.L  Ruzzo,  R.Smolensky and P. Tiwari, P. Proceedings of the Twenty First Annual ACM Symposium on Theory of Computing, Seattle, Washington, pp. 574–586 (1989)
\bibitem{randic} D. J. Klein and M. Randic, Journal of Mathematical Chemistry. 12, 81 (1993)
\bibitem {Garcia} A. Tizghadam and A. Leon-Garcia, IEEE Network 24, 10 (2010).
\bibitem{Biggs} N. Biggs, Algebraic Graph Theory, second edition,  Cambridge University Press, (1996)
\bibitem{Chau} C-K Chau, and P. Basu, IEEE/ACM TRANSACTIONS ON NETWORKING. 19,  4, ( 2011)
\bibitem{Lovasz} L.Lovasz, Random walk on graphs: A survey. combinatorics, 2 1 (1993)
\bibitem{wu} F. Y. Wu,  J. Phys. A: Math. Gen. 37 6653 (2004).
\bibitem{Aldous} D.J Aldous,  J.Fill, 2012. Reversible Markov chains and random walks on graphs. Available in: http://www.stat.berkeley.edu/~aldous/RWG/book.html
\bibitem{Schwatt} I. J. Schwatt, An Introduction to the Operations with Series. Philadelphia, press of the university of Pennsylvania, (1924).
\bibitem{Chair1} Noureddine Chair,  submitted
\bibitem{Wolf} K.B. Wolf, Integral transforms in science and engineering  New York and London Plenum press (1979)
\bibitem{Egorychev} G. P. Egorychev.Integral representation and the computation of combinatorial sums American Mathematical Soc, 1984.
\bibitem{Rivlin} T. J Rivlin. Chebyshev Polynomials. New York: Wiley,
 1990 
 \bibitem{Sloane} N. J. A. Sloane, The On-Line Encyclopedia of Integer Sequences, http://oeis.org
\bibitem{Stein} C.J. Everrett, P.R. Stein, One-dimensional random walk with absorbing barriers, Discrete Math. 17 27 (1977)
\bibitem{Chair} Noureddine Chair,  Journal of Geometry and Physics 38  170 2001
\end{thebibliography}

\end{document}